\documentclass[prb,twocolumn]{revtex4}%

\usepackage{color}
\usepackage{amsmath}
\usepackage{graphicx}
\usepackage{textcomp}
\usepackage{tabularx}
\usepackage{ulem}

\newcommand{\be}{\begin{equation}}
\newcommand{\ee}{\end{equation}}
\newcommand{\bea}{\begin{eqnarray*}}
\newcommand{\eea}{\end{eqnarray*}}
\newcommand{\bean}{\begin{eqnarray}}
\newcommand{\eean}{\end{eqnarray}}

\begin{document}

\draft
\title
{\bf Superlattice nanowire heat engines with direction-dependent
power output and heat current}

\author{ David M T Kuo$^{1}$, and Yia-Chung
Chang$^{2,3}$}

\address{$^{1}$Department of Electrical Engineering and Department of Physics, National Central
University, Chungli, 320 Taiwan}

\address{$^{2}$Research Center for Applied Sciences, Academic Sinica,
Taipei, 11529 Taiwan}

\affiliation{$^3$ Department of Physics, National Cheng Kung
University, Tainan, 701 Taiwan}

\date{\today}

\begin{abstract}
Heat engines (HEs) made of low dimensional structures offer
promising applications in energy harvesting due to their reduced
phonon thermal conductance. Many efforts have been devoted to the
design of HEs made of quantum-dot (QD) superlattice nanowire
(SLNW), but only SLNWs with uniform energy levels in QDs were
considered. Here we propose a HE made of SLNW with staircase-like
QD energy levels. It is demonstrated that the nonlinear Seebeck
effect can lead to significant electron transports for such a
nanowire with staircase-like energy levels. The asymmetrical
alignment of energy levels of quantum dots embedded in nanowires
can be controlled to allow resonant electron transport under
forward temperature bias, while they are in off- resonant regime
under backward bias. Under such a mechanism,the power output and
efficiency of such a SLNW are better than SLNWs with uniform QD
energy levels. The SLNW HE has direction-dependent power output
and heat current. In addition, the HE has the functionality of a
heat diode with impressive negative differential thermal
conductance under open circuit condition.
\end{abstract}

\maketitle
\textbf{1. Introduction}

Recently, many efforts were devoted to the studies of the
nonlinear thermoelectric properties of low dimension systems for
the applications of energy
harvesting.[\onlinecite{Kuo1}-\onlinecite{Pietzonka}] The figure
of merit ($ZT$) of quantum dots (QDs) junction system  may
approach infinity (corresponding to Carnot efficiency) in the weak
coupling between QD and electrodes when systems have a vanishingly
small phonon thermal conductance.[\onlinecite{Murphy}] However,
their electrical power output is extremely weak due to very small
electron tunneling rates.[\onlinecite{Kuo1},\onlinecite{Murphy}] A
remarkable thermoelectric device needs not only a high efficiency
but also significant power
output.[\onlinecite{Whitney1},\onlinecite{Whitney2}] Therefore,
how to design a heat engine with near Carnot efficiency and
optimized power output is under hot
pursuit.[\onlinecite{Hart}-\onlinecite{Pietzonka}]  QD
superlattice nanowires (SLNWs) offer high potential to realize
significantly reduced phonon thermal
conductance.[\onlinecite{Harm},\onlinecite{Zeb}] It is expected
that the efficiency of SLNW heat engines (HEs) is relatively high
when compared with other low dimensional systems. Nevertheless,
theoretical studies of SLNW HEs reported so far are based on the
assumption of uniform energy levels in QDs without considering the
effect of nonlinear Seebeck voltage resulting from temperature
bias.[\onlinecite{Whitney1},\onlinecite{Karbaschi},\onlinecite{Kuo2}]

Most recently, Ref.[\onlinecite{Craven}]has pointed out that
nonlinear Seebeck voltage plays a remarkable role for heat diode
design by considering hetero-molecular junctions. Therefore, we
attempt to reveal the effect of Seebeck voltage on the power
output and TE efficiency of SLNW HEs. Furthermore, we also
demonstrate that electron heat diodes can be implemented by using
nonlinear Seebeck voltage of a QD SLNW with staircase-like energy
levels. The design structure is shown in Fig. 1. Although the
staircase-like energy levels of QDs in a SLNW make it difficult
for the electron transport under a small temperature bias, a
suitable alignment of QD energy levels can be designed to allow
resonant electron transport under large forward temperature bias,
while the system is in off-resonant regime under backward bias.
This mechanism can give rise to a high efficiency and optimized
power output for SLNW HEs.

\textbf{2. Theoretical method}

To study the direction-dependent nonlinear thermoelectric
properties of QD SLNW connected to metallic electrodes shown in
Fig. 1(a), we start with the system Hamiltonian given by an
extended Anderson model
$H=H_0+H_{QD}$[\onlinecite{Haug},\onlinecite{Jauho}], where
\begin{eqnarray}
H_0& = &\sum_{k,\sigma} \epsilon_k
a^{\dagger}_{k,\sigma}a_{k,\sigma}+ \sum_{k,\sigma} \epsilon_k
b^{\dagger}_{k,\sigma}b_{k,\sigma}\\ \nonumber &+&\sum_{k,s}
V^L_{k,L}d^{\dagger}_{L,s}a_{k,\sigma}
+\sum_{k,s}V^R_{k,R}d^{\dagger}_{R,s}b_{k,\sigma}+c.c.
\end{eqnarray}
The first two terms of Eq.~(1) are for  free electrons in the left
and right electrodes. $a^{\dagger}_{k,\sigma}$
($b^{\dagger}_{k,\sigma}$) creates  an electron of momentum $k$
and spin $\sigma$ with energy $\epsilon_k$ in the left (right)
electrode. $V^L_{k,L}$ ($V^R_{k,R}$) describes the coupling
between the leftmost (rightmost) QD in the SLNW and the left
(right) electrode. $d^{\dagger}_{L(R),s}$ ($d_{L(R),s}$) creates
(destroys) an electron in the leftmost (rightmost) QD. $s$ labels
the degenerate states in a QD level with orbital and spin
degeneracy.[\onlinecite{Kuo3}] Silicon has six equivalent valleys,
each having an ellipsoidal shape. In a SLNW with strong lateral
confinement along x and y directions, the energy of electron
states in two valleys elongated along the z axis will be lifted up
more than electronic states in the remaining four valleys. Thus,
we consider the four-fold valley degeneracy for each QD.

\begin{eqnarray}
H_{QD}&=&\sum_{\ell,s} E_{\ell} n_{\ell,s}+\sum_{\ell \neq j}
t_{\ell,j} d^{\dagger}_{\ell,s} d_{j,s}+ c.c
\end{eqnarray}
where $E_{\ell}$ denotes the energy of the level of the $\ell$-th
QD, and $t_{\ell,j}$ describes the electron hopping strength
between the $\ell$-th QD and its nearest neighbor QD labeled  by
$j$. For the SLNW depicted in Fig. 1(a), $E_\ell$ depends on the
location of  QD. Here, we assume the QD energy levels  have a
staircase-like distribution as shown in Fig. 1(b) in which
$E_N=E_R$, and $E_{\ell}=E_R+(N-\ell)\Delta E$, where $\Delta E$
denotes the energy level separation. Such a variation in QD levels
can be engineered by considering suitable diameter variation of
QDs in the SLNW shown in Fig. 1(a), which can be realized by the
advanced etch and lithography
technique[\onlinecite{Harm},\onlinecite{Kuo2}]. The electron
Coulomb interactions have been neglected in {Eq.~(2).} The
electron Coulomb interactions are weak under the resonant
tunneling condition, since the electron wavefunction in SLNW
becomes delocalized.Although intra and inter-dot electron Coulomb
interactions are strong in the off-resonant condition, the effect
of electron Coulomb interactions on the electron transport becomes
small when the electron population of each QD is
small.[\onlinecite{Kuo3}]. This study is restricted in this
situation.

\begin{figure}[h]
\centering
\includegraphics[angle=-90,scale=0.3]{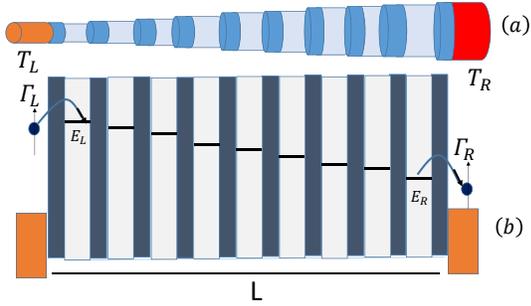}
\caption{(a) Schematic illustration of heat engine made of quantum
dots embedded in a nanowire with length $L$ connected to metallic
electrodes. (b) Energy diagram illustrating a QD array with
staircase-like alignment of energy levels connected to metallic
electrodes at $T_L=T_R$, where $T_L$ and $T_R$ are the equilibrium
temperature of left and right electrodes. $\Gamma_L$ ($\Gamma_R$)
denote the tunneling rates for electrons from the left (right)
electrode entering  the leftmost (rightmost) QD with energy level
$E_L$ ($E_R$).}
\end{figure}

The electron current from electrode to the QD SLNW can be derived
by using the Meir-Wingreen formula.[\onlinecite{Jauho}] We have
\begin{eqnarray}
J=\frac{se}{\hbar}\int \frac{d\epsilon}{2\pi}{\cal
T}_{LR}(\epsilon)[f_{L}(\epsilon)-f_R(\epsilon)],
\end{eqnarray}
where
$f_{\alpha}(\epsilon)=1/\{\exp[(\epsilon-\mu_{\alpha})/k_BT_{\alpha}]+1\}$
denotes the Fermi distribution function for the $\alpha$-th
electrode, where $\mu_\alpha$  and $T_{\alpha}$ are the chemical
potential and the temperature of the $\alpha$ electrode. $e$,
$\hbar$, and $k_B$ denote the electron charge, the Planck's
constant, and the Boltzmann constant, respectively. ${\cal
T}_{LR}(\epsilon)$ denotes the transmission coefficient of QD SLNW
connected to electrodes, which can be derived by the equation of
motion method.[\onlinecite{Kuo4}]

The expression of transmission coefficient is given by
\begin{equation}
{\cal T}_{L,R}(\epsilon)=\frac{4\Gamma_{L}(\epsilon)
\Gamma^{eff}_{R}(\epsilon)}{\Gamma_{L}(\epsilon)+\Gamma^{eff}_{R}(\epsilon)}
~(-Im(G^r_{L}(\epsilon))),
\end{equation}
where the tunneling rate
$\Gamma_{L(R)}(\epsilon)=2\pi\sum_{k}|V_{k,L(R)}|^2\delta(\epsilon-\epsilon_k)$.
In the wide band limit of electrodes, the energy-dependent
$\Gamma_{L(R)}(\epsilon)$ can be neglected. The notation Im means
taking the imaginary part of the function that follows, and
\begin{equation}
G^r_{L}(\epsilon)=1/(\epsilon-E_1+i\Gamma_L-\Sigma_{1,N})
\end{equation}
is the one-particle retarded Green function of the leftmost QD
with the energy level of $E_1$. The self energy
$\Sigma_{1,N}(\epsilon)$ results from electron tunneling from the
leftmost QD to the right electrode mediated by $N-1$ QDs, which is
given by [\onlinecite{Kuo2},\onlinecite{Kuo4}]

\be
\Sigma_{1,N}=\frac{t^2_{1,2}}{\epsilon-E_2-\frac{t^2_{2,3}}{\epsilon-E_3-
\cdots -\frac{t^2_{N-1,N}}{\epsilon-E_N+i\Gamma_R}, }} \ee where
$N$ denotes the total number of QDs. The rightmost QD is the $N$th
QD. The effective tunneling rate
$\Gamma^{eff}_R(\epsilon)=-Im(\Sigma_{1,N}(\epsilon))$. For
simplicity, we assume $t_{\ell,j}=t_c$ for all $\ell$ and $j$
being the nearest neighbor of $\ell$, and
$\Gamma_L=\Gamma_R=\Gamma$. Note that ${\cal T}_{LR}(\epsilon)$ of
QD chain can also be found in early work[\onlinecite{Teng}], where
authors calculated the Green's function in terms of matrix form.

The heat current for electrons leaving from the left (right) electrode is given by[\onlinecite{Kuo4}]

\begin{equation}
Q_{e,L(R)}=\frac{\pm s}{\hbar}\int \frac{d\epsilon}{2\pi}{\cal T}_{LR}(\epsilon)(\epsilon-\mu_{L(R)})[f_{L}(\epsilon)-f_R(\epsilon)].
\end{equation}
We note that $Q_{e,L}+Q_{e,R}=-(\mu_L-\mu_R) J/e$, which describes the Joule heating.

Because the phonon heat current $(Q_{ph})$ coexists with the
electron heat current, we should examine how $Q_{ph}$ will
influence the efficiency of a SLNW HE. To include $Q_{ph}$ we
adopted the following empirical formula given in
Ref.[\onlinecite{Chen}]

\begin{equation} Q_{ph}(T)= \frac{F_s}{\hbar} \int \frac{d\omega}{2\pi} {\cal T}(\omega)_{ph}
(\hbar^2 \omega)[n_{L}(\omega)-n_R(\omega)], \label{phC}
\end{equation}
where $\omega$ and ${\cal T}_{ph}(\omega)$ are the phonon
frequency and throughput function, respectively. ${\cal
T}_{ph}(\omega)$ depends on the length ($L$), surface roughness
width ($\delta$) and diameter ($D$) of silicon
nanowires.[\onlinecite{Chen}]
$n_{L(R)}=1/(exp(\hbar\omega/k_BT_{L(R)})-1)$. In the linear
response region, $Q_{ph}=\kappa_{ph}\Delta T$, where $\kappa_{ph}$
and $\Delta T$ are the phonon thermal conductance and temperature
difference between electrodes. In Ref.[\onlinecite{Chen}],
theoretical $\kappa_{ph}$ illustrates the experimental
$\kappa_{ph}$ of silicon nanowires below $300K$ very well. Because
authors have not considered the phonon-phonon collisions in
silicon nanowires, Eq. (8) is not adequate to describe $Q_{ph}$ in
high temperature region of $T > 300K$. A dimensionless factor
$F_s$ is introduced to describe the reduction factor for phonon
transport due to scattering from QDs embedded in a
nanowire.[\onlinecite{Hu}] The value of $F_s=0.1$ is used
throughout this article, which is determined according to
Ref.[\onlinecite{Hu}], in which the phonon thermal conductance of
silicon/germanium QD SLNWs is calculated. {Due to the low electron
density considered here and weak electron phonon interactions
(EPI) in Si/Ge, the electron mean free path ($\lambda$) of Si/Ge
QD SLNWs  is longer than $170~nm$ at room temperature. The length
of QD SLNW considered here is around $127~nm$, which is smaller
than $\lambda$ reported in Ref.[\onlinecite{Lu}]. Therefore, the
neglect of EPIs is justified.}

To design a heat engine driven by a high temperature-bias $\Delta
T=T_L-T_R$, the Seebeck voltage ($eV_{th}=\mu_L-\mu_R$) across the
external load with conductance $G_{ext}=1/R_{ext}$ needs to be
calculated.[\onlinecite{Josefsson}] Meanwhile, the energy levels
$E_{\ell}$ for all $\ell$ should be readjusted according to
$V_{th}$. As a consequence, ${\cal T}_{LR}(\epsilon)$ will depend
on $V_{th}$. The electron heat current satisfies the condition
{$Q_{e,L}+Q_{e,R}=-J V_{th}=P_{gen}$}, which denotes the work done
by the heat engine per unit time. The efficiency of heat engine is
defined as the power output divided by the power input. The power
input is the heat current out of the hot side and the power output
is the electrical power generated $P_{gen}$. Thus, the
direction-dependent efficiency of heat engine is given by
\begin{equation}
\eta_{\alpha}=- J V_{th}/|Q_{e,\alpha}|.
\end{equation}
We define $\Delta T > 0$ and  $\Delta T < 0$ as the forward
temperature bias and backward temperature bias. When $\Delta T> 0$
( $\Delta T< 0$), the electron heat current is leaving from the
left (right) electrode, $Q_{e,\alpha}$ in Eq. (9) denotes the
$Q_{e,L}$ ($Q_{e,R}$) in Eq. (7).

\textbf{3. Results and discussion}

We consider an $N=25$ SLNW with a staircase alignment of energy
levels. Namely, we have $E_L=E_1=E_R+24\Delta E$,
$E_2=E_R+23\Delta E$... and $E_N=E_R$. With an induced Seebeck
voltage, $V_{th}$, the energy levels $E_\ell$ are modified
according to $\varepsilon_\ell=E_\ell+\eta_\ell eV_{th}$. In a
simple approximation where the electric field is uniformly
distributed in spacer layers in the SLNW, the level modulation
factor is expressed as $\eta_{\ell}=-(\ell L_s-L/2)/L$ with
$\mu_{L(R)}=E_F\pm eV_{th}/2$. The pair length (that of one QD
plus one spacer layer) adopted is $L_s=5~nm$ and the length of
SLNW is $L=127 nm$.{\color{red}[\onlinecite{Hu}]} The Seebeck
voltage can be evaluated by Eq.~(3) under the condition of
$G_{ext}V_{th}+J(V_{th},\Delta T)=0$,  which is the same as the
experimental configuration in Ref [\onlinecite{Josefsson}], where
authors studied the HE made of a single QD connected to
electrodes. Once $V_{th}$ is obtained, the electron current $J$
and electron heat current {$Q_{e,L(R)}$} can be evaluated by
Eq.(3) and Eq.~(7), respectively. The resulting output power,
$P_{gen}$ and $V_{th}$ as functions of temperature bias for
various values of $\Delta E$ at
$t_c=\Gamma_{L}=\Gamma_{R}=1\Gamma_0$, {$G_{ext}=0.04G_0$} and
$E_R=E_F+4\Gamma_0$ are plotted in Fig.~2. {$G_0=e^2/h$  denotes
the quantum conductance  and $E_F$ is the Fermi energy of
electrodes. All energy scales are in units of $\Gamma_0$
throughout this article. The value of $\Gamma_0$ depends on the
desired temperature range considered in the design. In typical
designs considered, $\Gamma_0=1 meV$}

Figure~2(a) shows the asymmetrical behavior of output power,
$P_{gen}$.  It is found that $P_{gen}$  under backward temperature
bias ($\Delta T < 0$) is always smaller than that under forward
bias ($\Delta T > 0$). The asymmetry ratio,
$R_{asy}=P_{gen}(1\Gamma_0)/P_{gen}(-1\Gamma_0)$ is found to be 1,
1.43, and 2.81 for $\Delta E=0$, $0.05~\Gamma_0$, and
$0.1~\Gamma_0$, respectively. Such a ratio is enhanced with
increasing $\Delta T$. To understand the asymmetrical behavior of
$P_{gen}$, it is important to examine the relation between
$V_{th}$ and $\Delta T$. (See Fg. 2(b)) Nonlinear Seebeck
coefficients ($V_{th}/\Delta T$) are always negative, indicating
that electrons of the electrodes mainly diffuse through energy
levels above $E_F$. The Seebeck voltage not only changes chemical
potentials of electrodes ($\mu_{L(R)}=E_F\pm eV_{th}/2$), which
counter balances the electron flow from the hot side to the cold
side, but also influences the alignment of energy levels. With
forward temperature bias, the QD levels are tilted toward
alignment, allowing resonant tunneling of electrons from the left
electrode to the right electrode, while under reverse bias the QD
levels are further misaligned, leading to an off-resonance
condition. (See insets in Fig. 2(a)) In the limit of $\Delta T
\rightarrow 0$, we have $G_{ext}V_{th}+e^2{\cal L}_0 V_{th}+{\cal
L}_1\frac{\Delta T}{eT}=0$, where ${\cal L}_n=\frac{s}{h}\int
d\epsilon {\cal
T}_{LR}(\epsilon)(\epsilon-E_F)^n\frac{1}{4k_BT~cosh^2((\epsilon-E_F)/(2k_BT))}$
with $s=8$ resulting from the four-fold valley degeneracy and
spin. Note that transmission coefficient ${\cal T}_{LR}(\epsilon)$
in ${\cal L}_n$ is independent of $V_{th}$. The Seebeck voltage is
then given  by $V_{th}=\frac{-{\cal L}_1\Delta
T}{(eT)(G_{ext}+e^2{\cal L}_0)}$, which explains that $V_{th}$ and
$\Delta T$ always have opposite signs, if $E_R > E_F$ and
$P_{gen}$ is proportional to $\Delta T^2$. In the nonlinear
response region, ${\cal T}_{LR}(\epsilon)$ involves $V_{th}$, the
relation between $V_{th}$ and $\Delta T$ can be rather
complicated.

\begin{figure}[h]
\centering
\includegraphics[angle=-90,scale=0.3]{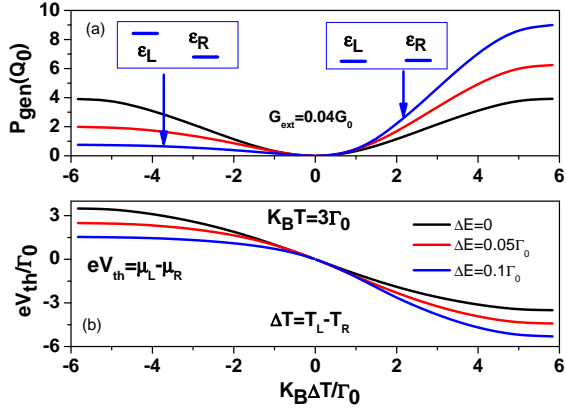}
\caption{(a) Electrical power output, (b) Seebeck-voltage as
functions of temperature bias for different $\Delta E$ values at
$T=36K$, $E_R=E_F+4\Gamma_0$, $t_{\ell,j}=t_c=1\Gamma_0$ and
$\Gamma_L=\Gamma_R=\Gamma=1\Gamma_0$. $Q_0=\Gamma^2_0/h$.
$T_L=T+\Delta T/2$ and $T_R=T-\Delta T/2$.}
\end{figure}

Because $P_{gen}(\Delta T > 0)$ is larger than $P_{gen}(\Delta T <
0)$ in Fig. 2(a), we further investigate the electron heat current
$Q_{e,L}$ and efficiency $\eta$ as functions of forward
temperature bias in Fig.~3(a) and 3(b), respectively. Like
$P_{gen}$, the electron heat current $Q_{e,L}$ at $\Delta E=0$ is
smaller than that at $\Delta E=0.1\Gamma_0$. In particular, the
maximum efficiency of HEs ($\eta_{max}$) at $\Delta E=0.1\Gamma_0$
is better than the case of $\Delta E=0$. The $\eta_{max}=0.75$ for
$\Delta E=0.1\Gamma_0$ corresponds to $ZT$ larger than fifteen
according to
$\eta/\eta_c=(\sqrt{ZT+1}-1)/(\sqrt{ZT+1}+1)$.[\onlinecite{Zeb}]
From the results of Fig. 2(a) and Fig.~3(b), we have demonstrated
that the $P_{gen}$ and $\eta$ of SLNW HEs with staircase-like
energy levels have better performance. Fig.~3(c) shows the
dependence of $\eta$ on the external load resistance, $R_{ext}$.
The $R_{ext}$-dependent $\eta$ has been investigated in the
experiment of Ref.[\onlinecite{Josefsson}] The dotted curve
includes the phonon heat current $Q_{ph}$, where a silicon
nanowire with surface roughness width $\delta=3~nm$, diameter
$D=3~nm$ and $L=127~nm$ is considered. The behavior of
$Q_{ph}(T,\Delta T$) was reported for different diameters of
silicon nanowires in our previous work [\onlinecite{Kuo6}]. The
suppression of $\eta=P_{gen}/(Q_{e,L}+Q_{ph})$ due to finite
$Q_{ph}$ is expected. The maximum $\eta$ occurs at $R_{ext}
\approx 20R_0$, {where $R_0=1/G_0$}. Note that when
$R_{ext}\rightarrow 0$, we have $V_{th}\rightarrow 0$, which leads
to vanishingly small $P_{gen}$. On the other hand, as
$R_{ext}\rightarrow \infty$ we have $J \rightarrow 0$ and
$P_{gen}\rightarrow 0$. Fig. 3(d) shows $\eta$ as a function of
$E_R$ at $\Delta E=0.1\Gamma_0$. The maximum $\eta$ occurs near
$E_R=E_F+5\Gamma_0$. In this case,  all QD energy levels are above
$E_F$, and the electron transport is mainly due to thermionic
process.

\begin{figure}[h]
\centering
\includegraphics[angle=-90,scale=0.3]{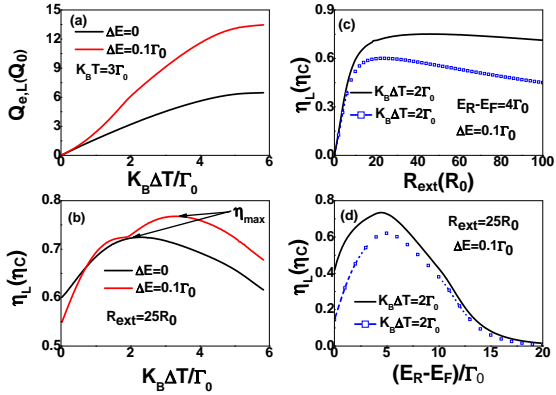}
\caption{(a) Electron heat current $Q_L$ and (b) efficiency
$\eta_L$ as functions of temperature bias for different $\Delta E$
values at $K_BT=3\Gamma_0$ and $E_R=E_F+4\Gamma_0$. (c) $\eta_L$
as functions of $R_{ext}$ at $K_B\Delta T=2\Gamma_0$, $\Delta
E=0.1\Gamma_0$, $K_BT=3\Gamma_0$ and $E_R=E_F+4\Gamma_0$. (d)
$\eta_L$ as functions of $E_R$ at $K_B\Delta T=2\Gamma_0$, $\Delta
E=0.1\Gamma_0$,$K_BT=3\Gamma_0$ and $R_{ext}=25R_0$. $R_0=1/G_0$.
Other physical parameters are
$t_c=\Gamma=1\Gamma_0$.$\eta_c=\Delta T/T_L$.}
\end{figure}

Heat diodes (HDs) play an important role in applications of energy
harvesting [\onlinecite{Terraneo}-\onlinecite{WangL}]. Those
designs considered three kinds of heat carriers:
phonons,[\onlinecite{Terraneo}-\onlinecite{Liy}], photons
[\onlinecite{Otey}], and electrons
[\onlinecite{Craven},\onlinecite{Kuo6}]. To investigate the
behavior of SLNW HDs, we consider the open-circuit condition of
$J=0$ [\onlinecite{Craven},\onlinecite{Kuo6}]  that
$Q_{e,L}=-Q_{e,R}=Q_e$ in which the contribution involving
$\mu_{L(R)}$ is zero in Eq. (7). The rectification ratio of HDs is
defined as $R_r = \frac{Q_{e}(\Delta T > 0)}{|Q_{e}(\Delta T <
0)|}$, where $Q_{e}(\Delta T>0) $ and $Q_{e}(\Delta T < 0)$ are
the heat currents in the forward and backward temperature bias,
respectively. Fig.~4(a) shows the calculated electron heat current
as a function of temperature bias, and the behavior of the
direction-dependent electron heat current (heat rectification) is
apparent. Note that $Q_e$ is not zero although $J=0$. Under
forward bias, a negative differential thermal conductance (NDTC)
is observed. To analyze the behavior of NDTC, we examine the
Seebeck voltage ($V_{th}$) as a function of $\Delta T$ {in Fig.
4(b).} For simplicity, let's consider $t_c=2~\Gamma_0$, which
corresponds to a narrow bandwidth case. For this case, the QD
energy levels are aligned (the resonant-tunneling condition) when
$K_B\Delta T=2.5~\Gamma_0$, which corresponds to
$eV_{th}=-10~\Gamma_0$ for $\Delta E=0.4~\Gamma_0$. When
$K_B\Delta T$ deviates from $2.5\Gamma_0$, the system is driven
away from the resonant condition. This explains why the electron
heat current has a peak near $K_B\Delta T=2.5\Gamma_0$, which
leads to NDTC as {$K_B\Delta T$ exceeds $2.5 \Gamma_0$.} In
Fig.~4(c), we show the electron heat rectification ratio ($R_r$)
as a function of $\Delta T$. Although the maximum $R_r$ reaches a
very high value near $60$ at $t_c=2~\Gamma_0$, the heat current is
very small. Good thermal diodes also require large heat current.
Thus, the cases with $t_c=3~\Gamma_0$ (red) and $4~\Gamma_0$
(blue) are better designs than the $t_c=2~\Gamma_0$ case, since
the heat current is significantly higher even though the maximum
$R_r$ is somewhat lower. The differential thermal conductances
(DTC) corresponding to the curves shown in Fig.~4(a) are given in
Fig. 4(d) in which very robust NDTC behavior is observed. The
feature of NDTC plays a remarkable role in the design of thermal
transistors.[\onlinecite{Hed}-\onlinecite{WangL}] So far, little
literature has reported NDTC resulting from electron carriers. It
is worth noting that the study of Ref.[\onlinecite{Craven}] does
not show the NDTC behavior for hetero-molecular junction, this may
be attributed to $\Delta T/T$ not large enough. According to the
results of Fig. 4, the ratio of $\Delta T/T > 1$ is preferred to
observe the feature of NDTC.

\begin{figure}[h]
\centering
\includegraphics[angle=-90,scale=0.3]{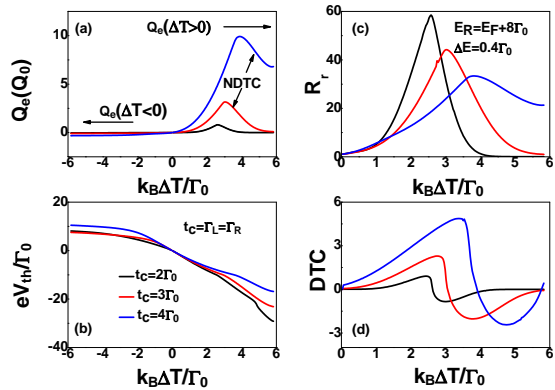}
\caption{(a)Electron heat current $Q_e$, (b) Seebeck voltage
$V_{th}$,(c) heat rectification ratio $\eta_R$ and (d)
differential thermal conductance (DTC) as functions of temperature
bias for different electron hopping strengths. $G_{ext}=0$,
$t_c=\Gamma_L=\Gamma_R$, $K_BT=3~\Gamma_0$ and
$E_R=E_F+8~\Gamma_0$.}
\end{figure}


\begin{figure}[h]
\centering
\includegraphics[angle=-90,scale=0.3]{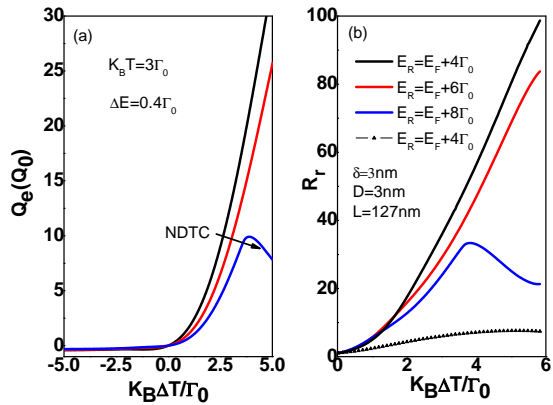}
\caption{(a)Electron heat current $Q_e$, and (b) heat
rectification ratio as functions of temperature bias for different
$E_R$ values at $t_c=\Gamma_L=\Gamma_R=4~\Gamma_0$,
$K_BT=3~\Gamma_0$, and $\Delta E=0.4~\Gamma_0$.}
\end{figure}

For further optimization we calculate $Q_e$ and $R_r$  as
functions of $\Delta T$ for various values of $E_R$ at
$t_c=\Gamma_L=\Gamma_R=4~\Gamma_0$,  $K_BT=3~\Gamma_0$, and
$\Delta E=0.4~\Gamma_0$. The results are shown in Fig.~5.  At a
given value of positive $\Delta T$, $Q_{e}(\Delta T> 0)$ is
suppressed with increasing $E_R$ due to the reduction of electron
population at high energy levels in the thermionic process. In the
thermal-assisted transport process, $V_{th}$ increases
significantly with increasing $E_R$ (not shown). In the cases of
$E_R=E_F+4~\Gamma_0$ and $E_R=E_F+6~\Gamma_0$, the magnitude of
$V_{th}$ is not enough to create the resonant-tunneling condition
for electron transport under forward bias. Therefore, no NDTC is
observed in these two cases. In Fig.~5(b), it is seen that
$\eta_R$ for $E_R=E_F+4~\Gamma_0$ reaches a very impressive value
of $100$. However, the maximum $\eta_R$ for this case is reduced
to near $10$ when $Q_{ph}$ is included (dotted curve). In the
design of semiconductor HDs with high $R_r$ values, obviously, the
reduction of $Q_{ph}$ is strongly required. It is expected that
the effect of phonon heat current can be reduced in the future
with the advances in nanotechnology [\onlinecite{Hu}]. Therefore,
the HEs and HDs of SLNW with staircase-like QD energy levels exist
the promising potential in the applications of energy harvesting.

4. Summary

In conclusion, we have theoretically investigated the
direction-dependent electrical power output and electron heat
rectification of a QD SLNW. The alignment of energy levels of QDs
in the nanowire can be altered by the temperature-bias induced
Seebeck voltage which leads to resonant tunneling for electrons in
forward bias but off-resonance in reverse bias for properly
designed distribution of QD energy levels in SLNW. This provides a
physical mechanism for achieving  direction-dependent $P_{gen}$
and $Q_e$. We found that the maximum efficiency and optimized
$P_{gen}$ of SLNW with staircase-like QD energy levels are better
than those of SLNW with uniform QD energy levels. In addition, we
have demonstrated that such SLNWs have unique behavior not only in
electron heat rectification but also in  NDTC, which is a key
ingredient for the implementation of thermal logical gates and
transistors.



\begin{flushleft}
{\bf Acknowledgments}\\
\end{flushleft}
This work was supported by the Ministry of Science and Technology (MOST) of Taiwan under contract nos. 107-2112-M-008-023-MY2 and 107-2112-M-001-032.

\mbox{}\\

$^{1}$E-mail address: mtkuo@ee.ncu.edu.tw\\
$^{2}$E-mail address: yiachang@gate.sinica.edu.tw\\

\setcounter{section}{0}

\renewcommand{\theequation}{\mbox{A.\arabic{equation}}} 
\setcounter{equation}{0} 

\mbox{}\\

\end{document}